\documentclass[a4paper,reqno,11pt]{amsart}

\usepackage[a4paper, scale={0.78,0.78}, marginratio={1:1}, footskip=7mm, headsep=10mm]{geometry}

\usepackage{amsmath,amsfonts,amssymb,amsthm,amscd,mathdots,bbm,graphicx,xcolor}
\usepackage[latin1]{inputenc}

\title[Combinatorial Exploration of Multidrug Polyspecificity in EPs]{Combinatorial Exploration of Multidrug Polyspecificity  in Efflux Pumps}

\author[F.~D.~Cunden]{Fabio Deelan Cunden}
\address{Dipartimento di Matematica, Univerisit\'a degli Studi di Bari, I-70125 Bari, Italy, and
INFN, Sezione di Bari, I-70126 Bari, Italy}
\email{fabio.cunden@uniba.it}

\author[J.~C.~Jimenez-Castellanos]{Juan Carlos Jimenez-Castellanos}
\address{Universidad Nacional Aut\'onoma de M\'exico, Departamento de Biolog\'ia. Laboratorio 1-A. Facultad de Qu\'imica, Edificio A, Cd. Universitaria, Coyoac\'an, 04510 Ciudad de M\'exico, Mexico}
\email{jcjcqfb@gmail.com}

\author[R.~Ortega Munoz]{Raquel Ortega Munoz}
\address{Universidad Nacional Aut\'onoma de M\'exico, Departamento de Biolog\'ia. Instituto de Ciencias Biom\'edicas. Cd. Universitaria, Coyoac\'an, 04510 Ciudad de M\'exico, Mexico}
\email{ortgm@quimica.unam.mx}

\begin{document}

\maketitle

\begin{abstract}
A defining feature of efflux pumps is multidrug polyspecificity, which to date still eludes some of the traditional dogmas of drug binding within protein science.
 We propose a combinatorial approach to explore the neighbourhood of efflux pump superfamilies in the vast sequence space of polypeptide chains. By generating new candidate structures through structured permutations of existing sequences, this framework aims to uncover hidden determinants of efflux pump functionality.
  
\end{abstract}

\subsection*{Phenomenology}

Efflux pumps (EPs), particularly those of the Resistance-Nodulation-Division (RND) superfamily (e.g., AcrB--TolC), play a central role in multidrug resistance (MDR) in clinically relevant Gram-negative bacteria such as \textit{Escherichia coli}, \textit{Klebsiella pneumoniae}, and \textit{Acinetobacter baumannii}. A hallmark of these systems is their ability to expel chemically diverse substrates (\emph{polyspecificity}), including many antibiotics~\cite{Du18}. 

 The binding and transport mechanisms of these substrates have been primarily studied in {efflux pumps} from \textit{E. coli} and \textit{K. pneumoniae} (i.e., AcrB), where both the sequences and structural similarities are high ($>90\%$). Significant questions, however, arise when considering how other pumps in Gram-negative bacteria (e.g., AdeB, AdeH, AdeJ, and MexB) bind and transport a wide range of similar substrates despite their low sequence similarity~\cite{JCJC23,Meier}. 

Our current understanding of the binding and transport of a wide range efflux pump substrates has primarily been based on traditional methods and using \textit{E. coli}'s AcrB pump as the role model. Substrate polyspecificity has typically been assessed one drug molecule at a time or by individually generating or selecting specific mutants~\cite{Save02}. While successful in specific cases, these methods provide limited insight into the global principles governing polyspecificity and hinder the systematic design of inhibitors or modified antibiotics capable of evading efflux.

We propose instead a combinatorial exploration of sequence space, generating hybrid proteins (EP-hybrids) from multiple RND pumps. This approach aims to reveal shared and hidden mechanisms of substrate transport and to provide a basis for rational inhibitor design.

\subsection*{Protein conformation and functionality}

Proteins are polymers composed of $20$ amino acids linked by peptide bonds, with lengths ranging from tens to thousands of residues. Interactions among amino acid side chains drive folding into a specific three-dimensional conformation.

In short, the folding of a polypeptide chain in a given environment is a solution as a large system of (possibly stochastic) dynamical equations, in which the final \emph{conformation} of the protein is the lowest free energy state: this is the  `uniqueness' property in Anfinsen's thermodynamic principle~\cite{Anfisen74}.  A direct solution to the folding problem is computationally unfeasible, and one usually resorts to comparing a given amino acid sequence to similar ones whose structure is already known by using matching and/or perturbation techniques. 
Although exact solutions remains challenging, major advances have recently been achieved using machine learning approaches~\cite{Repecka21,Hayes24,Jumper21}.

However, determining the folded structure is only a first step. The key objective in protein design is to identify the geometric and energetic features responsible for biological \emph{function} (e.g., binding or transport). We propose a combinatorial strategy to explore nearby regions of sequence space by generating structured variations of known proteins, complementing traditional approaches such as site-directed mutagenesis.

\subsection*{Combinatorial exploration}

Let $\mathcal{A}$ denote the set of all amino acid sequences over an alphabet of size $20$. A polypeptide chain is represented by a word $w \in \mathcal{A}$. The size of this space grows exponentially (e.g., $\sim 20^{1000}$ for typical EP lengths), while the subset $\mathcal{B} \subset \mathcal{A}$ corresponding to naturally occurring proteins is vanishingly small.

We assume a mapping
\[
w \mapsto \mathrm{Conf}(w),
\]
associating each sequence to its folded conformation (neglecting environmental variability for simplicity).

Functional features such as active sites are encoded by \emph{labels}, forming an abstract set $\mathbb{L}$. For a given label $\mathbb{L}$, we define its \emph{syntactic image} $L \subset \mathcal{A}$ as the set of sequences exhibiting that function (we translated this neat description from 
\cite{Carbone03}). Examples include sequences that fold correctly or exhibit specific transport capabilities.
 As a specific example, the RND superfamily can be thought of as the family of native sequences in $\mathcal{B}$ that contain (in the sense of substring inclusion) the syntactic image $L$ of the label $\mathbb{L}=\text{`ability to expel the chemical substrate $X$'}$. 

We propose the following combinatorial scheme:
\begin{enumerate}
\item Select a family of proteins sharing functional labels $\mathbb{L}_1, \mathbb{L}_2, \ldots$ (e.g., the RND superfamily);
\item Generate new sequences by applying permutations to each sequence in the family;
\item Analyse sequence and structural similarity using computational tools;
\item Experimentally assess the functionality of the resulting proteins.
\end{enumerate}

The central challenge lies in selecting meaningful permutations in step (2). Motivated by the observation that functional motifs vary across homologous proteins, we propose to preserve key substrings associated with functional labels while permuting the remaining sequence. This produces a family of sequences in which active sites are conserved but the surrounding context is altered.

We conjecture that many such permutations will create new families with functionalities $\mathbb{L}'_1,\mathbb{L}'_2,\ldots$, most likely the same $\mathbb{{L}}_1,\mathbb{{L}}_2,\ldots$ of the original family, indicating robustness of biological function to against generic perturbatio. 
of the `inactive sites'. The conjecture is supported by preliminary analysis of the sequences and structures similarity of permutations in the RND-superfamily. 

Conversely, it is reasonable to expect that some special permutations will split the family i.e., some polypeptide chains will have different functionality than the others (or will be not functional at all in some cases). 
Those singular
permutations would signal hidden mechanisms that participate in the functionality of 
proteins \emph{without} transforming or mutating the corresponding active sites.

\subsection*{Applications}

Protein combinatorics suggests new avenues for drug discovery. By integrating it with methods such as fluorescent assays, phenotypic profiling, Cryo-EM, and medicinal chemistry, it would be feasible to explore previously uncharted chemical spaces within efflux pumps, aiding to advance our structure-activity relationship 
 knowledge about EPs polyspecificity with the ultimate goal of designing new inhibitors and antibiotics that can evade efflux pumping.

\end{document}